\documentstyle[12pt]{article}
\newlength{\extraspace}
\newlength{\extraspaces}
\newcommand{\be}{\begin{equation}
\addtolength{\abovedisplayskip}{\extraspaces}
\addtolength{\belowdisplayskip}{\extraspaces}
\addtolength{\abovedisplayshortskip}{\extraspace}
\addtolength{\belowdisplayshortskip}{\extraspace}}
\newcommand{\ee}{\end{equation}}
\newcommand{\ba}{\begin{eqnarray}
\addtolength{\abovedisplayskip}{\extraspaces}
\addtolength{\belowdisplayskip}{\extraspaces}
\addtolength{\abovedisplayshortskip}{\extraspace}
\addtolength{\belowdisplayshortskip}{\extraspace}}
\newcommand{\ea}{\end{eqnarray}}
\newcommand{\newsection}[1]{
\vspace{7mm}
\pagebreak[3]
\addtocounter{section}{1}
\setcounter{equation}{0}
\setcounter{subsection}{0}
\setcounter{footnote}{0}
\begin{center}
{\large {\bf \thesection. #1}}
\end{center}
\nopagebreak
\medskip
\nopagebreak
\hspace{3mm}}
\newcommand{\nonu}{\nonumber \\[.5mm]}
\newcommand{\A}{&\!\!\!}

\begin{document}
\addtolength{\baselineskip}{.7mm}
\begin{flushright}
STUPP-95-139 \\ May, 1995
\end{flushright}
\vspace{.6cm}
\begin{center}
{\large{\bf{General considerations of matter coupling \\[2mm]
              with the self-dual connection}}} \\[20mm]
{\sc Motomu Tsuda, Takeshi Shirafuji and Hong-jun Xie} \\[12mm]
{\it Department of Physics, Saitama University \\[2mm]
Urawa, Saitama 338, Japan} \\[20mm]
{\bf Abstract}\\[10mm]
{\parbox{13cm}{\hspace{5mm}
It has been shown for low-spin fields that the use
of only the self-dual part of the connection
as basic variable does not lead to extra conditions
or inconsistencies.
We study whether this is true for more general
chiral action.
We generalize the chiral gravitational action,
and assume that half-integer spin fields are coupled
with torsion linearly.
The equation for torsion is solved and substituted
back into the generalized chiral action,
giving four-fermion contact terms.
If these contact terms are complex,
the imaginary part will give rise to extra conditions
for the gravitational and matter field equations.
We study the four-fermion contact terms
taking spin-1/2 and spin-3/2 fields as examples.}}
\end{center}
\vfill

\newpage
\setcounter{section}{0}
\setcounter{equation}{0}
\newsection{Introduction}

In the mid-1980s Ashtekar has presented a new formulation
of general relativity from a non-perturbative point of view,
in terms of which all the constraints of the gravity become
simple polynomials of complex canonical variables
\cite{AA}-\cite{AAL}.
It is possible to arrive at this formulation
from a first-order Palatini-type action
in which a tetrad and a complex self-dual connection
are regarded as independent variables \cite{AAL}.
Moreover, this formulation can be extended
to include matter sources.
In particular, it has been shown that the constraints
are again polynomials of the canonical variables
in the cases of spin-1/2 fields \cite{AAL,ART}
and $N = 1$ supergravity \cite{AAN,JJ,CDJ}.
The recent progress can be traced from the references
compiled in \cite{Bib.}.

It is the purpose of this paper to discuss
the consistency of the field equations
in the Ashtekar formalism.
For definiteness, we shall restrict ourselves
to the self-dual case throughout this paper.
The chiral action $S^{(+)}$ using the self-dual connection
can be expressed as \cite{AAL}
\ba
S^{(+)} \A = \A
[{\rm Einstein \! \! - \! \! Hilbert \ action}] \nonu
        \A \A + [{\rm quadratic \ terms \ of \ torsion}]
              + [{\rm matter \ terms}]. \label{action+2}
\ea
The total divergence terms of the action, if they arise,
are omitted throughout this paper,
because they do not contribute to the field equations.
Since the torsion vanishes in the source-free case,
the complex chiral action is equivalent
to the Einstein-Hilbert action.
If the matter terms (in particular, half-integer
spin fields) exist, however, the imaginary part
will emerge from the quadratic terms of torsion
and from the matter terms in (\ref{action+2}).
If this is indeed the case, the field equations
will be affected by the imaginary part of the chiral action.

In 1973 Hayashi and Bregman generalized the gravitational
Lagrangian of the Poincar\'{e} gauge theory
by adding the most general quadratic terms of torsion
to the first-order Palatini Lagrangian \cite{HB}.
It is easy to set up the chiral version
of their generalization.
When matter fields of half-integer spin exist,
the quadratic terms of torsion in the generalized
chiral action differ from those of (\ref{action+2}).
So it is of interest to study how the gravitational
and matter field equations are affected
by such a generalization.

As for the matter coupling, we shall obtain
the matter Lagrangian by the minimal prescription
which employs only the self-dual connection.
The Lagrangian of Dirac fields and Rarita-Schwinger fields
is, as is well-known, linear in first derivatives
of the field variables, and the torsion is linearly
coupled with these fields.
Therefore, we assume that the matter Lagrangian
for half-integer spin fields involves
the first order derivatives and hence the torsion linearly.

We are thus led to the generalized chiral action
which is schematically expressed as follows:
\ba
S^{(+)} \A = \A
[{\rm Einstein \! \! - \! \! Hilbert \ action}] + S_M(e) \nonu
        \A \A + [{\rm general \ quadratic \ terms \ of \ torsion}] \nonu
        \A \A + [({\rm torsion}) \times ({\rm matter \ fields})],
\label{action+4}
\ea
where $S_M(e)$ is the matter action in general relativity.
The equation of motion for the torsion
can be derived from (\ref{action+4}),
and can easily be solved with respect to the torsion.
Substituting the solution back into the torsion part
of (\ref{action+4}) gives the four-fermion contact terms
of the generalized chiral action.
If these contact terms are complex,
there is a possibility that consistency problem
arises in the gravitational and matter field equations.

It has been shown for spin-1/2 fields \cite{AAL,ART}
and simple supergravity \cite{AAN,JJ,CDJ} that the imaginary
part of the chiral action vanishes.
We investigate in this paper whether
the imaginary part of the four-fermion contact terms
vanishes or not for generic half-integer spin fields
which is assumed to couple with the torsion linearly.

For convenience of calculation, a tensor quantity
is introduced, which is denoted by $L_{ij \mu}$
and related to the torsion,
by taking the covariant differentiation of the quadratic
form of the tetrad field and by totally anti-symmetrizing
three space-time indices.
\footnote{\ Greek letters $\mu, \nu, \cdots$ are
space-time indices, and Latin letters {\it i, j,} $\cdots$
are local Lorentz indices.
We denote the Minkowski metric
by $\eta_{ij} =$ diag$(-1, +1, +1, +1)$.
The totally antisymmetric tensor $\epsilon_{ijkl}$
is normalized as $\epsilon_{0123} = +1$.}
This quantity will enable us to solve the equation
of motion for the self-dual part of the contorsion tensor
in a simple and transparent manner.

The present paper is organized as follows.
In Sec.2, following Hayashi and Bregman, we generalize
the chiral gravitational Lagrangian.
In Sec.3 we discuss the half-integer spin fields
minimally coupled to gravity.
Based on the considerations of spin-1/2 and spin-3/2 fields,
the general form of the chiral Lagrangian of matter fields
is set up assuming that the torsion is coupled
with the matter fields linearly.
In Sec.4 we derive the equation of motion for the torsion
from the generalized chiral Lagrangian,
and solve it with respect to the torsion.
In Sec.5 the consistency of the field equations is analyzed.
Spin-1/2 fields and (Majorana) Rarita-Schwinger fields
are considered as examples.
In the final section our results are summarized.

\newsection{Chiral gravitational Lagrangian and its generalization}

We consider space-time manifold with a metric
field $g_{\mu \nu}$ constructed from a tetrad field $e^i_{\mu}$
via $g_{\mu \nu} = e^i_{\mu} e^j_{\nu} \eta_{ij}$,
and denote a Lorentz connection by $A_{ij \mu} = A_{[ij] \mu}$.
\footnote{\ The antisymmetrization of a tensor is denoted
by $A_{[ij]} := (1/2)(A_{ij} - A_{ji})$.}
In the first-order Palatini framework, independent
dynamical variables are the tetrad and the Lorentz connection.
On the other hand, the tetrad and the complex self-dual connection
are used as the dynamical variables in the chiral action.
The self-dual connection $A^{(+)}_{ij \mu}$ is defined by
\be
A^{(+)}_{ij \mu} := {1 \over 2} \left(A_{ij \mu}
      - {i \over 2} {\epsilon_{ij}} \! ^{kl} A_{kl \mu} \right).
\ee
The complex chiral action using the tetrad
and the self-dual connection is
\be
S^{(+)} = \int d^4 x \, e R^{(+)} + [{\rm matter \ terms}], \label{action+1}
\ee
where $e = {\rm det}(e^i_{\mu})$ and
\be
R^{(+)} = {1 \over 2} \left({R^{ij}} \! _{\mu \nu}
                       - {i \over 2} {\epsilon^{ij}} \! _{kl}
            {R^{kl}} \! _{\mu \nu} \right) e^{\mu}_i e^{\nu}_j.
\label{R+}
\ee
Here ${R^{ij}} \! _{\mu \nu}$ is the curvature tensor
written in terms of the Lorentz connection,
and we use unit with $8 \pi G = c = 1 $.
The $R^{(+)}$ of (\ref{R+}) is also given by the self-dual
connection as
\be
R^{(+)} = (\partial_{\mu} {A^{(+)ij}} \! _{\nu}
                       - \partial_{\nu} {A^{(+)ij}} \! _{\mu}
  + {A^{(+)i}} \! _{k \mu} {A^{(+)kj}} \! _{\nu}
  - {A^{(+)i}} \! _{k \nu} {A^{(+)kj}} \! _{\mu}) e^{\mu}_i e^{\nu}_j.
\ee

In order to separate the quadratic terms of torsion
from the gravitational part of the chiral action
in a transparent manner, we introduce the following quantity:
\be
L^{ij \mu} := -{1 \over 4} {\epsilon^{ij}} \! _{mn}
         \epsilon^{\mu \nu \rho \sigma} D_{\nu} {H^{mn}} \! _{\rho \sigma},
                               \label{L-DH}
\ee
where ${H^{ij}} \! _{\mu \nu}$ is defined by
\be
{H^{ij}} \! _{\mu \nu} := 2e^i_{[\mu} e^j_{\nu]} \label{H-e}
\ee
and the covariant derivative $D_{\mu}$ acts
only on local Lorentz indices giving
\be
D_{\mu} e^i_{\nu} = \partial_{\mu} e^i_{\nu}
                        + {A^i} \! _{j \mu} e^j_{\nu}.
\label{cov.der.}
\ee
The Lorentz connection is divided into
the Ricci rotation coefficients $A_{ij \mu}(e)$
and contorsion tensor $K_{ij \mu}$:
\be
A_{ij \mu} = A_{ij \mu}(e) + K_{ij \mu}.
\label{A-Ae,K}
\ee
Using (\ref{A-Ae,K}) in (\ref{cov.der.}) gives
\be
D_{[\mu} e^i_{\nu]} = {K^i} \! _{j [\mu} e^j_{\nu]}
                    = -{1 \over 2} {T^i} \! _{\mu \nu}
\ee
with ${T^i} \! _{\mu \nu}$ being the torsion tensor.
Then the $L^{ij \mu}$ of (\ref{L-DH}) can be rewritten as
\be
L^{ij \mu} = -{1 \over 2} {\epsilon^{ij}} \! _{mn}
         \epsilon^{\mu \nu \rho \sigma} e^m_{\nu} {T^n} \! _{\rho \sigma},
                               \label{L-T}
\ee
which can be solved with respect to $T_{\mu ij}$ as follows:
\be
T_{\mu ij} = - L_{ij \mu} - e_{\mu [i} {L_{j]k}} \! ^k. \label{T-L}
\ee
Furthermore, $L_{ij \mu}$ can also be expressed
by the contorsion tensor as
\be
L_{ij \mu} = 2(e_{\mu [i} {K_{j]k}} \! ^k - K_{\mu [ij]}). \label{L-K}
\ee
With the help of this relation (\ref{L-K}),
the scalar curvature can be written as
\be
R = R(e) + {1 \over 2} L_{ij \mu} K^{ij \mu}, \label{R-LK}
\ee
where $R(e)$ is the Riemann-Christoffel scalar curvature.

The self-dual part of $L^{ij \mu}$ can be represented as
\be
L^{(+)ij \mu} = -{1 \over 4} {\epsilon^{ij}} \! _{mn}
      \epsilon^{\mu \nu \rho \sigma} D_{\nu} {H^{(+)mn}} \! _{\rho \sigma},
                         \label{L+-DH+}
\ee
where ${H^{(+)ij}} \! _{\mu \nu}$ is self-dual with respect to $(ij)$.
 From (\ref{L+-DH+}), we can obtain the following relation:
\be
L^{(+)}_{ij \mu} = 2(e_{\mu [i} {{K^{(+)}} \! _{j]k}} \! ^{k}
                          - K^{(+)}_{\mu [ij]}) \label{L+-K+}
\ee
with $K^{(+)}_{ij \mu}$ being self-dual with respect to $(ij)$.
Accordingly, the self-dual part of the scalar curvature
is written as
\be
R^{(+)} = {1 \over 2} R(e) + {1 \over 2} L^{(+)}_{ij \mu} K^{(+)ij \mu}
\label{R+-L+K+}
\ee
This expression of $R^{(+)}$ should be compared
to (\ref{R-LK}) for $R$.

We are now ready to separate the quadratic terms of torsion
from the gravitational part
of the chiral action of (\ref{action+1}).
Using (\ref{R+-L+K+}) in (\ref{action+1}),
the chiral action can be rewritten as follows:
\be
S^{(+)} = {1 \over 2} \int d^4 x \, e [R(e)
              + L^{(+)}_{ij \mu} K^{(+)ij \mu}]
              + [{\rm matter \ terms}]. \label{action+3}
\ee
In the source-free case, the torsion identically vanishes
and hence the chiral action of (\ref{action+3})
is reduced to the Einstein-Hilbert action.
If the matter terms exist, however, the imaginary part
appear in the chiral action of (\ref{action+3}).

We shall analyze the consistency of the field equations
which are derived from a slightly generalized chiral action.
Hayashi and Bregman proposed to extend the gravitational
Lagrangian in Poincar\'{e} gauge theory
by adding the most general kinetic Lagrangian $L_T$
of the tetrad field to the first-order Palatini Lagrangian:
\be
L_G = {1 \over 2} R + L_T.
\label{LG-R,LT}
\ee
According to the relation
$D_{[\mu} e^i_{\nu]} = {K^i} \! _{j [\mu} e^j_{\nu]}$,
the $L_T$ is a bilinear form of the torsion
(or equivalently the contorsion).
Thus the most general expression of $L_T$ is given by
\be
L_T = \alpha L_{ij \mu} K^{ij \mu}
                   + \beta v^2 + \gamma a^2 \label{LT-LK}
\ee
with $\alpha, \beta$ and $\gamma$
being three arbitrary parameters,
when invariance under space inversion is required.
\footnote{\ The $\alpha', \beta'$ and $\gamma'$ in eq.(4.2a)
of Ref.\cite{HB} are related to the $\alpha, \beta$ and $\gamma$
of (\ref{LT-LK}) by $\alpha' = (4/3)\alpha,
\beta' = \beta - (4/3)\alpha$ and $\gamma' = \gamma + 3 \alpha$.}
Here the vector $v_i$ and the axial vector $a_i$ are
\ba
v_i \A = \A {K_{ij}}^j = -{1 \over 2} {L_{ij}}^j, \\
a_i \A = \A {1 \over 3} \epsilon_{ijkl} K^{jkl}
    = -{1 \over 6} \epsilon_{ijkl} L^{jkl},
\ea
respectively, with $K_{ijk} = e_k^{\mu} K_{ij \mu}$
and $L_{ijk} = e_k^{\mu} L_{ij \mu}$.
The gravitational Lagrangian of (\ref{LG-R,LT})
can be rewritten as
\be
L_G = {1 \over 2} R(e)
      + \left[ \left(\alpha + {1 \over 4} \right) L_{ij \mu} K^{ij \mu}
               + \beta v^2 + \gamma a^2 \right],
\label{LG-LK,v,a}
\ee
by means of (\ref{R-LK}) and (\ref{LT-LK}).

The chiral version of this generalization
of the gravitational Lagrangian can be constructed
in the following manner.
The chiral kinetic Lagrangian
of the tetrad field, $L^{(+)}_T$, is obtained
by replacing the $L_{ij \mu}$ in (\ref{LT-LK})
by $L^{(+)}_{ij \mu}$; namely,
\be
L^{(+)}_T = 2(\alpha L^{(+)}_{ij \mu} K^{(+)ij \mu}
            + \beta v^{(+)2} + \gamma a^{(+)2}), \label{LT+-L+K+}
\ee
where
\ba
v^{(+)}_i \A = \A {{K^{(+)}}_{ij}}^j
          = -{1 \over 2} {{L^{(+)}}_{ij}}^j, \\
a^{(+)}_i \A = \A {1 \over 3} \epsilon_{ijkl} K^{(+)jkl}
    = -{1 \over 6} \epsilon_{ijkl} L^{(+)jkl}
\ea
with $K^{(+)}_{ijk} = e_k^{\mu} K^{(+)}_{ij \mu}$
and $L^{(+)}_{ijk} = e_k^{\mu} L^{(+)}_{ij \mu}$.
We note that since $K^{(+)}_{ijk}$ is self-dual
with respect to $(ij)$, $a^{(+)}_i$ and $v^{(+)}_i$
are related each other as follows:
\be
a^{(+)}_i = {2 \over 3}i v^{(+)}_i
          = {1 \over 2} \left(a_i + {2 \over 3}i v_i \right).
\label{a+-v+}
\ee
The chiral action of (\ref{action+1}) is now generalized
to the following form:
\be
S^{(+)} = \int d^4 x \, e L^{(+)}_G + [{\rm matter \ terms}]
\label{action+5}
\ee
with the generalized chiral gravitational Lagrangian
$L^{(+)}_G$ being
\be
L^{(+)}_G = R^{(+)} + L^{(+)}_T.
\label{LG+-R+,LT+}
\ee
With the help of (\ref{a+-v+}) and (\ref{R+-L+K+}),
$L^{(+)}_G$ can be rewritten as
\be
L^{(+)}_G = {1 \over 2} R(e)
           + 2 \left[ \left(\alpha + {1 \over 4} \right)
                     L^{(+)}_{ij \mu} K^{(+)ij \mu}
     + \left(\beta - {4 \over 9}\gamma \right) v^{(+)2} \right].
\label{LG+-LK+,v+}
\ee
It should be noticed that this involves two parameters
in contrast to the $L_G$ of (\ref{LG-LK,v,a})
with three parameters.

\newsection{Chiral Lagrangian of matter fields}

The Lagrangian of integer spin fields does not
contain the Lorentz connection and its imaginary part
vanishes as in the source-free case.
Therefore, we restrict ourselves to half-integer spin fields,
taking spin-1/2 and spin-3/2 fields as examples.

We start with two postulates for the matter coupling.
Firstly, let us suppose that the matter Lagrangian is obtained
by the minimal prescription; namely, by replacing ordinary
derivatives acting on half-integer spin fields
by covariant derivatives,
\be
\partial_i \rightarrow e_i^\mu D_\mu
\ee
with
\be
D_\mu = \partial_\mu + {i \over 2} A_{ij \mu} S^{ij},
\ee
and $S_{ij}$ standing for the SL(2,{\bf C}) generator.
\footnote{\ In our convention $S_{ij} = {i \over 4}[\gamma_i, \gamma_j]$
and $\{ \gamma_i, \gamma_j \} = - 2 \eta_{ij}$.}

Secondly, we suppose that the chiral Lagrangian
of half-integer spin fields
is described by using the self-dual part of the Lorentz connection.
According to the equation
\be
A^{(+)}_{ij \mu} S^{ij} = A_{ij \mu} S^{ij} {1 + \gamma_5 \over 2},
\ee
this demands that only the terms expressed by $D_\mu \psi_R$
and/or $\overline \psi_L \overleftarrow D_\mu$ should appear
in the matter Lagrangian. Here $\psi_R$ ($\psi_L$) is
the right (left)-handed spinor field:
\ba
\left\{ \matrix{\psi_R \A = \A \displaystyle {{1 + \gamma_5 \over 2}}
    \psi, \vspace{1mm} \cr
       \psi_L \A = \A \displaystyle {{1 - \gamma_5 \over 2}}
    \psi. \cr } \right.
\ea

We shall now see how the above postulates apply
to the matter coupling.
For example, let us consider
a (Majorana) Rarita-Schwinger field $\psi_{\mu}$.
Spin-1/2 fields can be treated in a similar manner.

The Lagrangian of a (Majorana) Rarita-Schwinger field
in flat space is usually defined by
\be
L^{(0)}_{RS} = {1 \over 2} \epsilon^{\mu \nu \rho \sigma}
              \overline \psi_\mu \gamma_5 \gamma_\rho
              \partial_\sigma \psi_\nu. \label{LRS-1}
\ee
If we apply the minimal prescription to (\ref{LRS-1}),
the expression of (\ref{LRS-1}) becomes
\be
L_{RS} = {1 \over 2} \epsilon^{\mu \nu \rho \sigma}
               \overline \psi_\mu \gamma_5 \gamma_\rho
               D_\sigma \psi_\nu, \label{LRSE}
\ee
which is used in $N = 1$ supergravity \cite{FNF}-\cite{NWH}.

According to the second postulate, however,
the Lagrangian (\ref{LRS-1}) in flat space
should be reexpressed only in terms of
$\partial_\mu \psi_{R \nu}$
and/or $\overline \psi_{L \nu} \overleftarrow \partial_\mu$.
In fact, the Lagrangian (\ref{LRS-1}) can be rewritten as
\be
L^{(0)}_{RS} = - \epsilon^{\mu \nu \rho \sigma} \overline \psi_{R \mu}
              \gamma_{\rho} \partial_{\sigma} \psi_{R \nu},
\label{LRS-2}
\ee
except for a total divergence term.
Note that only the first derivative of $\psi_{R \nu}$
appears in (\ref{LRS-2}),
because $\psi_{\nu}$ is Majorana spinor.
We can take (\ref{LRS-2}) as the chiral Lagrangian in flat space,
for which to apply the minimal prescription.
Then the chiral Lagrangian of a (Majorana) Rarita-Schwinger field
in curved space is given by
\be
L^{(+)}_{RS} = - \epsilon^{\mu \nu \rho \sigma}
                     \overline \psi_{R \mu} \gamma_\rho
                     D_\sigma \psi_{R \nu}. \label{LRSE+}
\ee

In order to see the relation between the $L_{RS}$
of (\ref{LRSE}) and the $L^{(+)}_{RS}$ of (\ref{LRSE+}),
we rewrite these Lagrangians
by using the contorsion tensor and its self-dual part.
Then the Lagrangians of (\ref{LRSE}) and (\ref{LRSE+})
can respectively be reexpressed as
\ba
L_{RS} \A = \A L_{RS}(e) + {i \over 4}
                  \epsilon^{\mu \nu \rho \sigma} \overline \psi_\mu
                  \gamma_5 \gamma_\rho K_{ij \sigma} S^{ij} \psi_\nu,
                  \label{LRSK} \\
L^{(+)}_{RS} \A = \A L_{RS}(e) + {i \over 2}
                  \epsilon^{\mu \nu \rho \sigma} \overline \psi_\mu
                  \gamma_5 \gamma_\rho K^{(+)}_{ij \sigma} S^{ij} \psi_\nu,
                  \label{LRSK+}
\ea
where $L_{RS}(e)$ is the Lagrangian
of a (Majorana) Rarita-Schwinger field in general relativity.
Comparing (\ref{LRSK+}) with (\ref{LRSK}),
we notice that the factor of the second term
in (\ref{LRSK+}) is twice that in (\ref{LRSK}),
and the real part of (\ref{LRSK+}) coincides
with that of (\ref{LRSK}).

We can treat a Dirac field $\psi$ similarly.
The Lagrangian of a Dirac field in curved space
can be obtained by applying the minimal prescription
to that in flat space.
On the other hand, in conformity with the second postulate,
the chiral Lagrangian of a Dirac field
in flat space can be defined by
the Lagrangian of a Dirac field
which is rewritten only
in terms of $\partial_\mu \psi_R$
and $\overline \psi_L \overleftarrow \partial_\mu$
except for a total divergence term.
We also apply the minimal prescription
to this chiral Lagrangian.
Furthermore, in order to see the relation between
the Lagrangian in curved space and the chiral one,
we use the contorsion tensor and its self-dual part.
The resulting Lagrangians, $L_D$ and $L^{(+)}_D$,
are given by
\ba
L_D \A = \A L_D(e)
       - {1 \over 4} \overline \psi \{ \gamma^{\mu}, S^{ij} \}
            K_{ij \mu} \psi, \label{LDK} \\
L^{(+)}_D \A = \A L_D(e)
       - {1 \over 2} \overline \psi \{ \gamma^{\mu}, S^{ij} \}
            K^{(+)}_{ij \mu} \psi, \label{LDK+}
\ea
where $L_D(e)$ is the Lagrangian
of the Dirac field in general relativity.
It can be seen that the factor of the second term
in (\ref{LDK}) is twice that in (\ref{LDK+}),
and the real part of (\ref{LDK+}) coincides
with that of (\ref{LDK})
as in the case of a (Majorana) Rarita-Schwinger field.

In the above two examples of a (Majorana) Rarita-Schwinger field
and a Dirac field, the Lagrangian
is linear in first derivatives of the field variables,
and the torsion is coupled with these fields linearly.
Thus let us suppose that the {\it real} Lagrangian
of half-integer spin fields is written as
\be
L_M = L_M(e) + X_{ij \mu} K^{ij \mu}, \label{LMK}
\ee
where $L_M(e)$ is the Lagrangian
of half-integer spin fields in general relativity,
and $X_{ijk} = X_{[ij] k}$ is a real tensor made of matter fields.
Then, taking into account of the above comparison the real
Lagrangian with the complex chiral one,
we assume that the {\it complex} chiral Lagrangian
of half-integer spin fields is expressed as follows:
\be
L^{(+)}_M = L_M(e) + 2X_{ij \mu} K^{(+) ij \mu}.
\label{LMK+}
\ee
It is noted that only the self-dual part $X^{(+)}_{ij \mu}$
contributes in (\ref{LMK+}).

\newsection{Solving with respect to the torsion}

The total chiral Lagrangian $L^{(+)}$
is given by the sum of (\ref{LG+-LK+,v+})
and (\ref{LMK+}):
\ba
L^{(+)} \A = \A L^{(+)}_G + L^{(+)}_M \nonu
        \A = \A {1 \over 2} R(e) + L_M (e)
                      + L^{(+)}_{{\rm torsion}},
\label{totL+}
\ea
where $L^{(+)}_{{\rm torsion}}$ is defined by
\be
L^{(+)}_{{\rm torsion}} = 2(\alpha' L^{(+)}_{ijk} K^{(+)ijk}
                  + \beta' v^{(+)2} + X^{(+)}_{ijk} K^{(+)ijk})
\label{torsion-L+K+}
\ee
with $\alpha'$ and $\beta'$ being set as
\be
\alpha' = \alpha + {1 \over 4}, \ \
\beta' = \beta - {4 \over 9}\gamma.
\ee
The imaginary part of the chiral Lagrangian $L^{(+)}$
emerges only from the $L^{(+)}_{{\rm torsion}}$
of (\ref{torsion-L+K+}).
The Lagrangian of (\ref{torsion-L+K+}) can be divided
into real and imaginary parts as follows:
\ba
L^{(+)}_{{\rm torsion}} \A = \A \alpha' L_{ijk} K^{ijk}
      + {1 \over 2} \beta' \left(v^2 - {9 \over 4}a^2 \right)
      + X_{ijk} K^{ijk} \nonu
      \A \A + i \left( -{\alpha' \over 2}
                      \epsilon_{ijmn} L^{ijk} {K^{mn}} \! _k
    -{3 \over 2} \beta' v_i a^i
    -{1 \over 2} \epsilon_{ijmn} X^{ijk} {K^{mn}} \! _k \right).
\label{torsion-Re,Im}
\ea

On the other hand, the total real Lagrangian $L$,
which is defined by the sum of (\ref{LG-LK,v,a})
and (\ref{LMK}), is
\be
L = L_G + L_M
  = {1 \over 2} R(e) + L_M (e)
              + L_{{\rm torsion}}
\label{L-LG,LM}
\ee
with $L_{{\rm torsion}}$ being defined by
\be
L_{{\rm torsion}} = \alpha' L_{ijk} K^{ijk}
                + \beta v^2 + \gamma a^2 + X_{ijk} K^{ijk}.
\ee
Comparing $L_{{\rm torsion}}$
and Re$L^{(+)}_{{\rm torsion}}$ of (\ref{torsion-Re,Im}),
we see that Re$L^{(+)}$ does not coincide with
$L$ generally unless $\beta = -(4/9)\gamma$.

Now it is easy to solve the equation of the torsion.
The results can be summarized as follows:

(1) If $K_{ij \mu}$ is taken as independent
variable instead of $A_{ij \mu}$, it can be shown that
the equation for $K_{ij \mu}$ derived
from the Re$L^{(+)}_{{\rm torsion}}$
of (\ref{torsion-Re,Im}) is equivalent to
that derived from Im$L^{(+)}_{{\rm torsion}}$
of (\ref{torsion-Re,Im});
namely, although $L^{(+)}$ is complex, extra conditions
do not appear for the equation for $K_{ij \mu}$.

(2) It can also be shown that the equation for $K^{(+)}_{ij \mu}$
obtained from the $L^{(+)}_{{\rm torsion}}$
of (\ref{torsion-L+K+})
is compatible with the equation for $K_{ij \mu}$
derived from the $L^{(+)}_{{\rm torsion}}$
of (\ref{torsion-Re,Im}).

(3) Varying the $L^{(+)}_{{\rm torsion}}$
of (\ref{torsion-L+K+}) with respect to $K^{(+)}_{ij \mu}$,
we get the following equation:
\be
2 \alpha' L^{(+)}_{ijk}
+ {1 \over 2} \beta' \left(\eta_{k[i} {{L^{(+)}} \! _{j]l}}^l
- {i \over 2} {\epsilon_{ijk}}^l {{L^{(+)}} \! _{lm}} \! ^m \right)
         = - X^{(+)}_{ijk}. \label{LLL+-X+}
\ee
This equation can be solved with respect to $L^{(+)}_{ij \mu}$ as
\be
L^{(+)}_{ijk} = -{1 \over 2 \alpha'}
        \left[ X^{(+)}_{ijk} - {f \over 2}
                     \left(\eta_{k[i} {{X^{(+)}}_{j]l}}^l
               - {i \over 2} {\epsilon_{ijk}} \! ^m
                   {{X^{(+)}}_{ml}}^l \right) \right],
\label{L+-XXX+}
\ee
where
\be
f := {\beta' \over
        {2 \alpha' - \displaystyle {3 \over 4} \beta'}}.
\ee
 From (\ref{L+-XXX+}) one can derive $L_{ij \mu}$.
Then the contorsion tensor $K_{ij \mu}$ can be
obtained by using the relation of (\ref{L-K}).
Substituting these $L_{ij \mu}$ and $K_{ij \mu}$
back into (\ref{torsion-Re,Im}),
we see that Re$L^{(+)}_{{\rm torsion}}$
and Im$L^{(+)}_{{\rm torsion}}$ give
four-fermion contact terms.

\newsection{Consistency of the field equations}

In order to get the explicit expression for Im$L^{(+)}$,
it is more convenient to write Im$L^{(+)}$
in terms of the torsion tensor.
Using $X_{ijk}$, which can be read
from (\ref{LLL+-X+}), in the Im$L^{(+)}_{{\rm torsion}}$
of (\ref{torsion-Re,Im}) gives
\be
{\rm Im}L^{(+)} = {\alpha' \over 2} \epsilon_{ijmn}
           L^{ijk} {K^{mn}} \! _k + {3 \over 2} \beta' v_i a^i.
\label{ImLK}
\ee
After a little calculation, (\ref{ImLK}) becomes
\be
{\rm Im}L^{(+)} = {\alpha' \over 2} \epsilon_{ijmn}
            T^{kij} {T_k}{\! ^{mn}} + {3 \over 2} \beta' v_i a^i.
\label{ImLK-T}
\ee

Let us now consider the cases of spin-1/2 and spin-3/2 fields.
In the case of spin-1/2 fields, the torsion tensor
is totally anti-symmetric:
\be
T_{kij} = -{1 \over 8 \alpha'} \left(1 + {3 \over 4} f \right)
              \epsilon_{ijkl} \overline \psi
                   \gamma_5 \gamma^l \psi. \label{TT-1/2}
\ee
Substituting (\ref{TT-1/2}) into (\ref{ImLK-T}),
we see that Im$L^{(+)}$ vanishes
for arbitrary values of $\alpha'$ and $\beta'$.

Next for $N$-(Majorana) Rarita-Schwinger fields $\psi^I_{\mu}$
where $I$ runs from 1 to $N$, the torsion tensor becomes
\be
T_{kij} = -{i \over 8 \alpha'}
     \sum_I \left[\overline \psi_i^I \gamma_k \psi_j^I
        - {f \over 2} \eta_{k[i} \overline \psi_{j]}^I \gamma^m \psi_m^I
        + {f \over 8} \epsilon_{ijkl} \epsilon^{lmnr}
                    \overline \psi_m^I \gamma_r \psi_n^I \right].
\label{TT-RS}
\ee
Inserting this into (\ref{ImLK-T}) gives
\be
{\rm Im}L^{(+)} = -{1 \over 128 \alpha'}
                    \sum_{I,J} \epsilon^{\mu \nu \rho \sigma}
            [(\overline \psi_{\mu}^I \gamma_k \psi_{\nu}^I)
             (\overline \psi_{\rho}^J \gamma^k \psi_{\sigma}^J)
    - f e_{\rho}^j e^{\lambda}_k
             (\overline \psi_{\mu}^I \gamma_j \psi_{\nu}^I)
             (\overline \psi_{\sigma}^J \gamma^k \psi_{\lambda}^J)].
\label{ImL-RS}
\ee
In this case Im$L^{(+)}$ does not vanish in general.

The gravitational and matter field
equations are affected by this Im$L^{(+)}$ of (\ref{ImL-RS}).
If the tetrad is real,
the Einstein equation and the Rarita-Schwinger equation
are obtained by taking variation of Re$S^{(+)}$
with respect to the tetrad and
the (Majorana) Rarita-Schwinger fields.
Furthermore, Im$S^{(+)}$ yields algebraic equations:
\be
\sum_{I,J}
f \epsilon^{\mu \nu \rho \sigma}
                  (\overline \psi_{\mu}^I \gamma_j \psi_{\nu}^I)
      [e^k_{\rho} e^j_{\lambda} e_i^{\tau}
                  (\overline \psi_{\sigma}^J \gamma^i \psi_{\tau}^J)
     - e^j_{\rho} (\overline \psi_{\sigma}^J \gamma^k \psi_{\lambda}^J)] = 0
\label{RS-4}
\ee
for the tetrad field $e^{\lambda}_k$, and
\be
\sum_{J(J \not= I)}
\epsilon^{\mu \nu \rho \sigma}
       [\overline \psi_{\rho}^J \gamma^k \psi_{\sigma}^J
  - f e^k_{\rho} e_j^{\lambda}
       (\overline \psi_{\sigma}^J \gamma^j \psi_{\lambda}^J
    + 2 \overline \psi_{\sigma}^I \gamma^j \psi_{\lambda}^I)]
                            \gamma_k \psi_{\nu}^I = 0
\label{RS-3}
\ee
for a (Majorana) Rarita-Schwinger field $\psi_{\mu}^I$.
In the derivation of (\ref{RS-3}),
we have used the identity
\be
\epsilon^{\mu \nu \rho \sigma}
    (\overline \psi_{\rho}^I \gamma^k \psi_{\sigma}^I)
   \gamma_k \psi_{\nu}^I \equiv 0, \label{Fierz}
\ee
which is valid because of the Fierz identity.
\footnote{\ See, for example, eq.(7) in p.365 of Ref.\cite{NWH}.}
When $f \not= 0$ (namely, $v^{(+) 2}$-term in (\ref{torsion-L+K+})
is non-vanishing), there is the difficulty of overdetermination
for $\psi_{\mu}^I$ from (\ref{RS-4}) and (\ref{RS-3}).
On the other hand, when $f = 0$
(namely, $v^{(+) 2}$-term in (\ref{torsion-L+K+})
is vanishing),
there is no such problem for $N = 1$
because both (\ref{RS-4}) and (\ref{RS-3}) disappear.
However, for $N \ge 2$, there is the possible difficulty
of overdetermination arising from (\ref{RS-3}).
It should be noticed that these results
are also valid for (Dirac) Rarita-Schwinger fields,
because a complex (Dirac) Rarita-Schwinger field
is equivalent with the two real (Majorana)
Rarita-Schwinger fields.

\newsection{Summary}

We studied the consistency of the field equations
by using the generalized chiral Lagrangian $L^{(+)}$
of (\ref{totL+}).
Let us summarize our results in order.

[1] The generalized gravitational Lagrangian $L_G$
involves three parameters $\alpha, \beta$ and $\gamma$
besides the Newton gravitational constant,
while its chiral part $L^{(+)}_G$ has two parameters
$\alpha' = \alpha + 1/4$ and $\beta' = \beta - (4/9)\gamma$.
The Re$L^{(+)}_G$ does not coincide with
the $L_G$ generally unless $\beta = -(4/9)\gamma$.

[2] The equation of motion for the contortion $K_{ij \mu}$
derived from Re$L^{(+)}$ is equivalent to
that derived from Im$L^{(+)}$; namely, although
$L^{(+)}$ is complex, extra conditions do not
appear for the equation of motion for $K_{ij \mu}$.

[3] The equation of motion for $K^{(+)}_{ij \mu}$
derived from $L^{(+)}$ is compatible with
the equation of motion for $K_{ij \mu}$
derived from $L^{(+)}$.

[4] Substituting $L_{ij \mu}$ obtained from the equation
of motion for $K^{(+)}_{ij \mu}$ back into $L^{(+)}$,
the torsion part of $L^{(+)}$ gives
the four-fermion contact terms, of which
the imaginary part is given by (\ref{ImLK-T}).

As examples, the cases of spin-1/2 fields
and (Majorana) Rarita-Schwinger fields
were discussed with the following results.

[5] For spin-1/2 fields,
the imaginary part of the generalized chiral Lagrangian
is always vanishing
for arbitrary values of $\alpha'$ and $\beta'$.
Therefore the consistency problem does not arise
in the case of spin-1/2 fields.

[6] For $N$-(Majorana) Rarita-Schwinger fields,
Im$L^{(+)}$ is non-vanishing generally.

(a) In the case of $\alpha = \beta = \gamma = 0$
(i.e. $L_G = (1/2)R$ and $L^{(+)}_G = R^{(+)}$),
the imaginary part of the chiral Lagrangian
does not vanish except for $N = 1$.
Thus the consistency problem arises for $N \ge 2$.

(b) In the general case of arbitrary values of $\alpha, \beta$
and $\gamma$, the imaginary part of the generalized
chiral Lagrangian does not vanish
unless the two conditions $f = 0$
(namely, $v^{(+) 2}$-term in (\ref{torsion-L+K+})
is vanishing) and $N = 1$ are satisfied.
If $f \not= 0$, therefore, the consistency problem arises.
When $N \ge 2$, there is such problem even if $f = 0$.

These results (a) and (b) are also valid
for (Dirac) Rarita-Schwinger fields.



\end{document}